\documentclass[sigconf]{acmart}

\AtBeginDocument{%
  }

\setcopyright{acmlicensed}
\copyrightyear{2025}
\acmYear{2025}
\acmDOI{XXXXXXX.XXXXXXX}

\acmConference[Conference acronym 'XX]{Make sure to enter the correct
  conference title from your rights confirmation email}{June 03--05,
  2018}{Woodstock, NY}
\acmISBN{978-1-4503-XXXX-X/18/06}
\usepackage{pifont}

\usepackage{amsmath,amsfonts}
\usepackage{array}
\usepackage{textcomp}
\usepackage{stfloats}
\usepackage{url}
\usepackage{verbatim}
\usepackage{graphicx}
\usepackage{pifont}
\usepackage[ruled,vlined]{algorithm2e}
\usepackage[justification=centering]{caption}
\usepackage{enumitem}
\usepackage{subcaption}
\usepackage{booktabs}
\usepackage{makecell}
\usepackage{tcolorbox}
\usepackage{soul}
\usepackage{ragged2e}
\usepackage{threeparttable}
\usepackage{balance}
\usepackage{float}
\usepackage{multirow}
\usepackage{arydshln}
\usepackage{arydshln}
\usepackage{xcolor}
\usepackage{tcolorbox}
\usepackage{graphicx}  
\usepackage{subcaption} 
\usepackage{adjustbox}
\newcommand{\find}[1]{
\begin{tcolorbox}[leftrule=0.5mm,rightrule=0.5mm, toprule=0.5mm,bottomrule=0.5mm,left=2pt,right=2pt,top=2pt,bottom=2pt]
\em #1
\end{tcolorbox}
}

\newcommand{\code}[1]{\texttt{#1}}
\newcommand{\approach}{{\textsc{CodeReasoner}}\xspace}

\begin{document}

\title{\approach: Enhancing the Code Reasoning Ability with Reinforcement Learning}

\author{Lingxiao Tang}
\authornote{Also with Hangzhou High-Tech Zone (Binjiang) Institute of Blockchain and Data Security}
\email{12421037@zju.edu.cn}
\orcid{0009-0003-7406-7961}
\affiliation{%
  \department{The State Key Laboratory of Blockchain and Data Security}
  \institution{Zhejiang University}
  \city{Hangzhou}
  \state{Zhejiang}
  \country{China}
}

\author{He Ye}
\email{he.ye@ucl.ac.uk}
\orcid{0000-0003-4807-2110}
\affiliation{%
  \institution{University College London}
  \city{London}
  \country{United Kingdom}}

\author{Zhongxin Liu}
\email{liu_zx@zju.edu.cn}
\orcid{0000-0002-1981-1626}
\affiliation{%
 \department{The State Key Laboratory of Blockchain and Data Security}
  \institution{Zhejiang University}
  \city{Hangzhou}
  \state{Zhejiang}
  \country{China}
}

\author{Xiaoxue Ren}
\email{xxren@zju.edu.cn}
\orcid{0000-0002-5526-1617}
\affiliation{%
 \department{The State Key Laboratory of Blockchain and Data Security}
  \institution{Zhejiang University}
  \city{Hangzhou}
  \state{Zhejiang}
  \country{China}
}

\author{Lingfeng Bao}
\authornote{Corresponding author}
\authornotemark[1]
\email{lingfengbao@zju.edu.cn}
\orcid{0000-0003-1846-0921}
\affiliation{%
    \department{The State Key Laboratory of Blockchain and Data Security}
  \institution{Zhejiang University}
  \city{Hangzhou}
  \state{Zhejiang}
  \country{China}
}

\renewcommand{\shortauthors}{Lingxiao Tang, He Ye, Zhongxin Liu, Xiaoxue Ren, and Lingfeng Bao}

\begin{abstract}
Code reasoning is a fundamental capability for large language models (LLMs) in the code domain. It involves understanding and predicting a program’s execution behavior across multiple dimensions, such as identifying the output given a specific input or determining whether a particular statement will be executed. This ability is critical for enhancing the performance of downstream tasks such as debugging, code generation, and program repair.
Previous approaches have primarily relied on supervised fine-tuning to improve LLMs' performance in code reasoning tasks. However, these methods often exhibit limited performance improvements and struggle to generalize across diverse reasoning scenarios.
We argue that this stems from two fundamental issues: the poor quality of existing training data and the inherent limitations of supervised fine-tuning, which often fails to teach models to generalize across diverse reasoning scenarios.
To address these limitations, we propose \approach—a novel framework that spans both dataset construction and a two-stage training process. First, we introduce a dataset construction method that focuses on capturing the core execution logic of Python programs. We then apply instruction tuning to inject execution-specific knowledge distilled from a powerful teacher model into the base LLM. Finally, we enhance the model’s reasoning ability and generalization through GRPO reinforcement learning applied on top of the fine-tuned model.

Extensive evaluation on three widely-used code reasoning benchmarks shows that \approach achieves performance improvements ranging from 27.1\% to 40.2\% over prior methods, when applied to a 7B-sized model. Remarkably, this 7B model achieves performance comparable to GPT-4o on key tasks such as input/output prediction and coverage prediction. When scaled to a 14B model, \approach outperforms leading models like GPT-4o across all datasets on average.
Ablation studies confirm the effectiveness of each training stage in \approach, and further analysis highlights the critical role of the reasoning chains in enhancing code reasoning performance.
\end{abstract}



\received{20 February 2025}  
\received[revised]{12 March 2025} 
\received[accepted]{5 June 2025}  

\maketitle

\section{Introduction} \label{sec:introduction}
Code reasoning ability is crucial for large language models (LLMs) because it enables LLMs to understand and predict the behavior of programs during execution. This capability is particularly important for tasks like debugging~\cite{zhong2024debug} and program repair~\cite{ni2024next,ye2022neural}, which require accurate simulation of code execution and correct understanding of the control and data flow.

To evaluate the LLM's code reasoning abilities, existing research has introduced a variety of tasks and benchmarks. These benchmarks, such as CRUXEval~\cite{gu2024cruxeval} and REval~\cite{chen2024reasoning}, are designed to test models on code reasoning tasks. These include predicting the output based on given inputs, inferring inputs from outputs, and answering detailed questions, such as whether a particular line of code will be executed. 
Compared to benchmarks like HumanEval~\cite{chen2021evaluating} and LiveCodeBench~\cite{jain2024livecodebench}, which have been widely used for evaluating code generation, code reasoning benchmarks offer a complementary perspective. They provide a deeper look into the LLM's understanding of program execution.

Experimental results on recent benchmarks reveal two key problems regarding the code reasoning capabilities of LLMs. First, there is a clear coding task bias between code generation and reasoning tasks~\cite{xu2024cruxeval,gu2024cruxeval}. Models that perform well on code generation tasks often struggle with code reasoning, indicating that the ability to generate correct code does not necessarily reflect a true understanding of the program behavior.
Second, there is a noticeable performance gap between smaller open-source models and larger models in code reasoning tasks. 
In code generation, 7B-sized models (e.g., Qwen2.5-Coder-Instruction) can achieve strong results, reaching a pass@1 score of 84.1\%\cite{hui2024qwen2} on the dataset HumanEvalPlus\cite{liu2023your}, just a few percentage points behind larger models like GPT-4o~\cite{GPT-4o}.
However, on reasoning-focused benchmarks like CRUXEval~\cite{gu2024cruxeval}, the performance gap widens significantly, reaching up to 20 percentage points (see Section~\ref{sec:Experiment results}).

Researchers have proposed several methods to improve the code reasoning abilities of LLMs. Two representative examples are SEMCODER~\cite{ding2024semcoder} and CODEI/O~\cite{li2025codei}. Both approaches follow a similar two-step pipeline. First, they use a teacher model to generate chains of thought for input/output prediction tasks. These reasoning traces are collected using rejection sampling~\cite{casella2004generalized} to ensure correctness. Then, the student model is fine-tuned on the distilled reasoning data to learn the teacher’s thought process.
While these methods have led to noticeable progress, several important limitations remain. Most notably, their performance still lags far behind that of more advanced models. In addition, these models often struggle to generalize to more fine-grained code reasoning tasks, such as predicting whether a specific line of code will be executed, resulting in significant performance degradation across benchmarks (see Section~\ref{sec:Experiment results}). 
We attribute this to two key issues.
First, the quality of current training data is low. Our investigation reveals that existing datasets often include excessive boilerplate code unrelated to core execution logic, which hinders effective learning.
Second, supervised fine-tuning (SFT) has inherent limitations. It struggles to generalize across tasks~\cite{gupta2025selective,wang2022two}, and recent studies show that SFT models can unexpectedly fail when presented with simple variations of their training data~\cite{lampinen2025generalization}.

To address the limitations outlined above, we propose \approach, a comprehensive framework that includes both dataset construction and a two-stage training process. We begin by introducing a novel dataset construction method designed to capture the core execution logic of code, avoiding irrelevant boilerplate and focusing on what truly matters for reasoning.
Building on the assumption that small models lack the reasoning patterns required to simulate the program behavior (see Section~\ref{subsec:preliminary1}), we first apply instruction tuning~\cite{peng2023instruction} to inject execution-specific knowledge distilled from a teacher model.
However, instruction tuning alone can lead to issues such as overly long reasoning chains and repetitive outputs~\cite{fu2021theoretical,dong2025rethinking}, which undermine both clarity and performance. To mitigate this and improve generalization, we introduce the GRPO reinforcement learning algorithm~\cite{guo2025deepseek} in the second training stage. GRPO encourages the model to generate more concise, accurate reasoning, leading to better performance across a wide range of code reasoning tasks.

We evaluate \approach across a diverse set of benchmarks, covering tasks from input/output prediction~\cite{gu2024cruxeval,jain2024livecodebench} to more fine-grained code reasoning challenges~\cite{chen2024reasoning}. 
Experimental results show that \approach significantly outperforms existing baselines and, in many tasks, narrows the gap with advanced models like GPT-4o when applied to a 7B-sized model. When scaled to 14B, \approach surpasses GPT-4o across all datasets on average, demonstrating its effectiveness and strong generalization ability.
We further conduct ablation studies to validate the contribution of each training stage in our framework and demonstrate the critical role of reasoning chains in performance improvement. Additionally, we analyze the GRPO training process in the context of code reasoning and compare this to its training process in other domains, such as mathematics~\cite{liu2025understanding,xie2025logic,zeng2025simplerl}. This comparison offers insights into why certain training dynamics, such as response length, exhibit different trends across domains.

In summary, we make the following contributions:
\begin{itemize}[leftmargin=*]
\item We propose a new framework named \approach to improve the code reasoning ability in small-sized LLMs, which includes training dataset construction and a two-phase training process. To the best of our knowledge, this is the first work to directly incorporate reinforcement learning into code reasoning tasks.
\item We evaluate \approach on a wide range of benchmarks, covering different code reasoning tasks. Experimental results show that \approach significantly outperforms existing baselines and achieves comparable performance to advanced models in many tasks, demonstrating its effectiveness and generalizability.
\item We conduct ablation studies to validate the contribution of each training stage and prove the effectiveness of the reasoning chains in performance improvement.
\item We also analyze the GRPO training process in the code reasoning domain and compare it to other domains (e.g., mathematics), offering insights into why certain training dynamics differ across tasks.
\item To support future research, we have publicly released our model and code, which are available online\footnote{\url{https://github.com/lingxiaotang/CodeReasoner}}.
\end{itemize}

\section{Preliminary Study} 
\label{sec:preliminary}
\begin{figure}[t]
  \centering
  \includegraphics[width=1.00\linewidth]{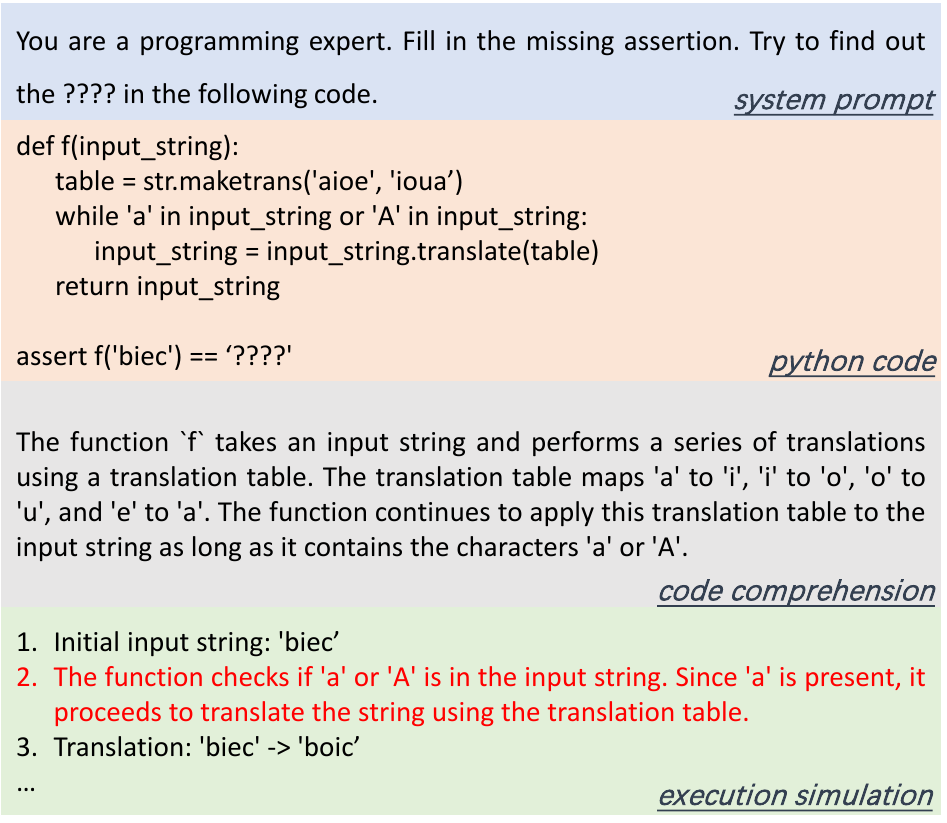}
  \caption{Motivating example from the Cruxeval benchmark showing the LLM correctly comprehends code but fails during execution simulation.}\label{fig:motivation1}
\end{figure}


In this section, we first explore why small-sized LLMs struggle with code reasoning tasks. We then analyze the limitations of existing training datasets designed to enhance LLMs’ code reasoning capabilities and highlight areas for improvement.

\subsection{Investigating LLM Failures in Code Reasoning}\label{subsec:preliminary1}

In this study, we first aim to investigate the strengths and weaknesses of large language models (LLMs) in understanding code execution. Specifically, we evaluate the Qwen2.5-7B Instruct model on the Cruxeval benchmark~\cite{gu2024cruxeval}.
 For each test case, the model is provided with a Python program and a corresponding input to its entry point, then asked to predict the program's expected execution output.
We randomly select 50 failed test cases and analyze them manually to identify common error patterns and underlying causes of failure.

The reasoning process of the considered LLM typically consists of two phases: \ding{182} code comprehension and \ding{183} execution simulation.
In the code comprehension phase, the model forms a logical blueprint of the program’s structure and intent without actual execution. This involves identifying the high-level purpose of each code block, decomposing fundamental operations and data structures, and inferring control flow (e.g. conditionals, loops and function boundaries).
In the execution simulation phase, the model simulates the program behavior step-by-step, initializing the program state using the given input, updating variable states after each operation, and precisely following the control flow, including conditional branches and loops. The predicted output corresponds to the final state after simulating the complete execution.

\textbf{The majority (47/50) failures occur during execution simulation.} Only three failure cases are related to code comprehension. We observe that the LLM struggles to accurately track complex string or list operations. Figure~\ref{fig:motivation1} illustrates a typical failure: the model correctly comprehends the creation of a translation table and its conditional application based on characters 'a' or 'A' (as shown in the middle part). 
However, during execution simulation, the model mistakenly assumes the input 'biec' contains 'a' or 'A', erroneously enters the loop, and outputs 'boic' instead of 'biec' (highlighted in red in the last part).

This limitation arises because LLMs simulate code execution based on statistical patterns learned from training data, and they are not actual code interpreters. As a result, they are prone to errors in tasks that require precise tracking of program states. We attribute this gap to the nature of the training process, which is primarily focused on static code–natural language pairs and lacks exposure to dynamic execution traces. This explains why such models tend to perform well in code comprehension or generation tasks but struggle with execution simulation—an observation consistent with findings from prior studies~\cite{gu2024cruxeval,xu2024cruxeval}.

\begin{figure}[t]
  \centering
  \includegraphics[width=0.95\linewidth]{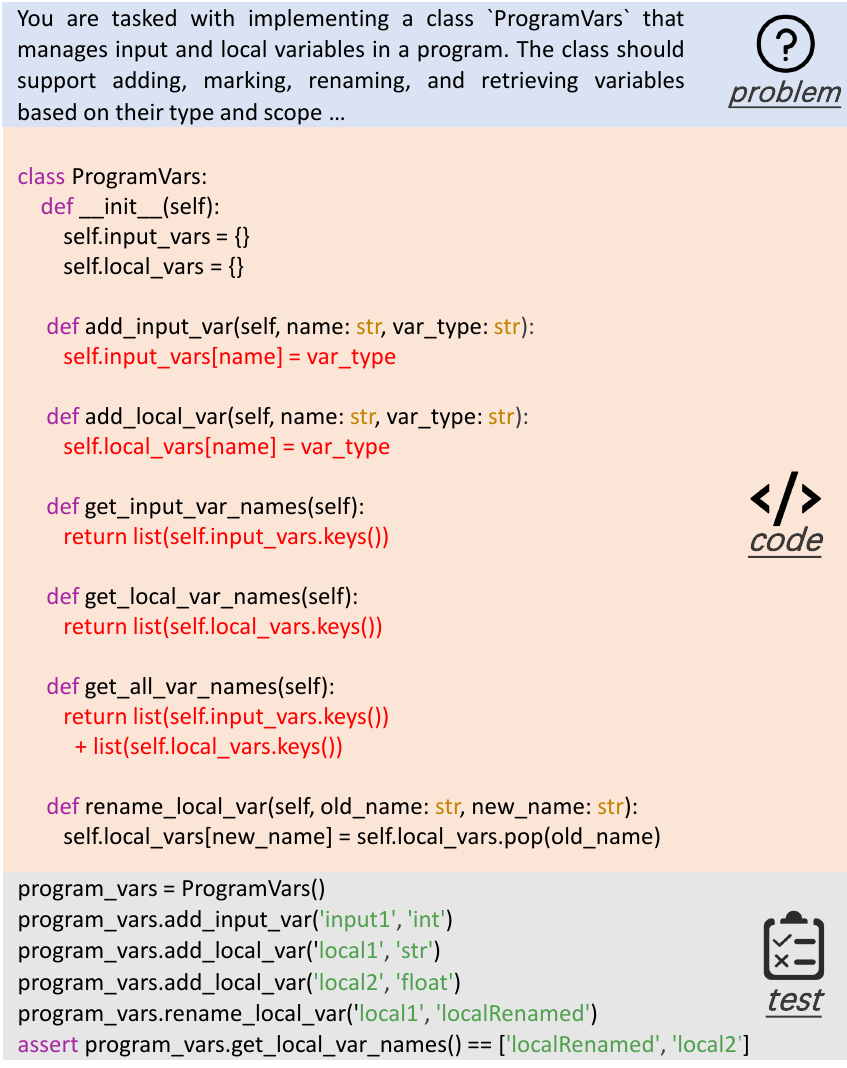}
  \caption{Motivation example from the PXY-R dataset. Boilerplate code (in red) adds unnecessary complexity.}\label{fig:motivation2}
\end{figure}

\subsection{Limitations of Prior Reasoning Datasets}\label{subsec:preliminary2}
Previous datasets aimed at enhancing the LLM's execution simulation, such as SEMCODER~\cite{ding2024semcoder} and CODEI/O~\cite{li2025codei}, are typically built through a three-step process. 
First, a code snippet is extracted from a real-world project to serve as a seed. 
Second, a large language model (LLM) is prompted to use this snippet as a basis to formulate a programming problem and generate a complete, functional program to solve it. Finally, an LLM or a dedicated input generator (e.g.,~\cite{liu2023your}) produces a suitable input for the program's entry point function, which completes the full data instance.

We identify two main limitations in datasets constructed this way. First, since LLMs are already strong at code comprehension (as shown in Section~\ref{subsec:preliminary1}), asking them to generate natural language descriptions for code is often unnecessary. 
Second, more critically, the data generation process tends to produce examples that appear complex but are simple to execute. When prompted to create programming problems, LLMs often imitate the structure of real-world software by adding classes, comments, and documentation, even when the core logic is very simple. This limits the effectiveness of such data for training models to understand the actual execution behavior.

We give a typical example in Figure~\ref{fig:motivation2} to illustrate this issue, selected from the PXY-R dataset by Ding et al.~\cite{ding2024semcoder}. 
Due to space limitations, we present its three key components: a) the problem statement; b) source code; and c) test case. 
The code defines a class with numerous boilerplate getter and setter functions, highlighted in red. Although the structure implies complexity, the core logic is simple. It simply tests basic method calls on Python's built-in data structures, such as dictionaries and lists, without involving any complex control flow.

We observe that existing code reasoning datasets often prioritize realistic code structure over actual reasoning difficulty, which leads to weak training signals for execution simulation. Much of the code does little to help LLMs improve in simulating program behavior. Repeated boilerplate functions obscure the simple underlying logic, causing the model to focus on code structure rather than reasoning. This motivates our work: built-in methods and control flow can be tested more effectively using simpler examples that remove extra code and highlight core execution logic. A more targeted approach to dataset construction is clearly needed.

\section{Approach} \label{sec:approach}

\begin{algorithm}[t!] 
\small
  \DontPrintSemicolon
  \SetKwData{builtinTypes}{builtinTypes}
  \SetKwData{builtinType}{builtinType}
  \SetKwData{method}{method}
  \SetKwData{useNestedCalls}{useNestedCalls}
  \SetKwData{useOtherMethods}{useOtherMethods}
  \SetKwData{controlFlows}{controlFlows}
  \SetKwData{dataset}{dataset}
  \SetKwData{baseCase}{baseCase}
  \SetKwData{mutatedCases}{mutatedCases}
  \SetKwData{testCase}{testCase}
  
  \SetKwFunction{getMethods}{getMethods}
  \SetKwFunction{randomBool}{randomBool}
  \SetKwFunction{getControlStmts}{getControlStmts}
  \SetKwFunction{llm}{llm\_generate} 
  \SetKwFunction{mutate}{mutate}
  \SetKwFunction{isValid}{isValid}
  
  \SetKw{Add}{Add}
  \SetKw{To}{to}
  \SetKw{From}{from}
  \SetKw{Remove}{Remove}

  \SetKwInOut{Output}{Output}
  
  \Output{\dataset, the final generated dataset.}
  
  \nl$\dataset \longleftarrow \emptyset$\;
  
  \tcp{Phase 1: Test Case Generation}
  \nl\For{\builtinType \textbf{in} \builtinTypes}{
    \nl\For{\method \textbf{in} \builtinType.\getMethods()}{
        \tcp*[l]{Whether to involve nested calls of the method in the test case}
        \nl$\useNestedCalls \longleftarrow \randomBool()$
        
        \tcp*[l]{Whether to involve other methods in the test case}
        \nl$\useOtherMethods \longleftarrow \randomBool()$ 
        
        \tcp*[l]{Get random control flow structures}
        \nl$\controlFlows \longleftarrow \getControlStmts()$ 
        
        \tcp*[l]{Use LLM to generate a base test case based on constraints}
        \nl$\baseCase \longleftarrow \llm(\method, \useNestedCalls,$\\ 
        \qquad $\useOtherMethods,\controlFlows)$ 
        
        \nl$\mutatedCases \longleftarrow \mutate(\baseCase)$\;
        
        \nl\Add \baseCase \To \dataset\;
        \nl\Add all cases from $\mutatedCases$ \To \dataset\;
    }
  }

  \tcp{Phase 2: Validation and Filtering}
  \nl\For{\testCase \textbf{in} \dataset}{
    \nl\If{not \isValid(\testCase)}{
        \nl\Remove \testCase \From \dataset\;
    }
  }
  
  \nl\textbf{return} \dataset \;
  \caption{Algorithm for Dataset Construction}
  \label{alg:test-suite-generation}

\end{algorithm}

\begin{figure*}[t]
  \centering
  \includegraphics[width=0.80\textwidth]{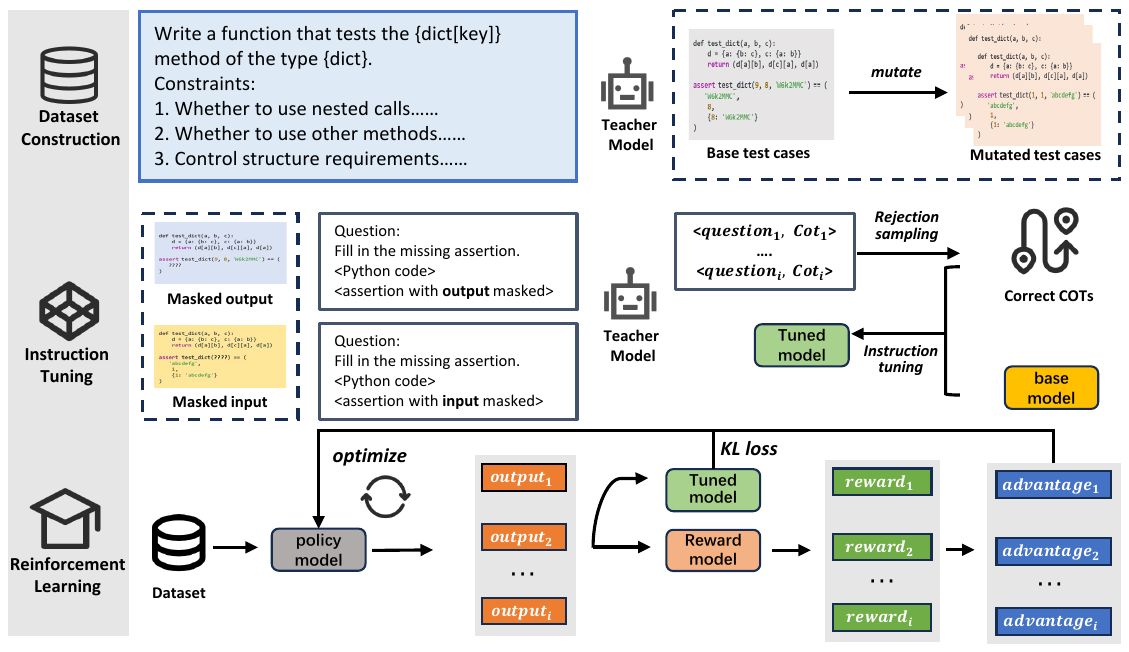}
  \caption{Overview of  \approach}\label{fig:overview}
\end{figure*}

Figure~\ref{fig:overview} presents the core ideas of \approach, which consists of three main stages. 
First, we construct a high-quality training dataset composed of concise cases specifically designed to enhance the model's code execution reasoning (as shown in the top part). 
Second,  we perform instruction tuning~\cite{peng2023instruction}, using this dataset to distill chain-of-thought (CoT) reasoning paths from a powerful teacher model (middle part). 
Finally, We refine the model using reinforcement learning to reduce overly long or repetitive CoT generations and improve its generalization ability (bottom part).
Now we discuss each part below.

\subsection{Dataset Construction} ~\label{subsection:Dataset Construction}
We aim to build a dataset with the following four key characteristics:
\ding{182} \textbf{Concise}: Each test case should be succinct and free of redundant elements such as boilerplate functions or wrappers. This ensures the dataset focuses on the core execution logic.
\ding{183} \textbf{Comprehensive}: The dataset should cover a wide range of execution scenarios that a model might encounter in practice.
\ding{184} \textbf{Controllable}: The generation process must be highly controllable. Rather than relying on unconstrained LLM generation, we introduce specific constraints to guide code creation.
\ding{185} \textbf{Varied Difficulty}: The dataset must contain test cases spanning various difficulty levels. Cases that are either too simple or overly complex can hinder effective training, particularly during the reinforcement learning phase.

Algorithm~\ref{alg:test-suite-generation} illustrates the two-phase pipeline used to construct our dataset.
In Phase 1 (Test Case Generation), the algorithm iterates over all built-in types and their associated methods in Python.
In Phase 2 (Validation and Filtering), each test case is executed and discarded if it fails validation.
We now describe the \textit{Method Call Constraints}, \textit{Control Structure Constraints}, and \textit{Mutation-based Augmentation} in Phase 1, as well as the \textit{Test Filtering} in Phase 2.

\textbf{Method Call Constraints:} We control the complexity of method interactions using several configuration parameters, as shown in Lines 4–5 of Algorithm~\ref{alg:test-suite-generation}. Specifically, the $useNestedCalls$ flag (Line 4) determines whether the LLM should generate nested method calls for the base method. Likewise, the $useOtherMethods$  flag (Line 5) directs the LLM to incorporate one or more additional methods into the generated code, creating more complex interactions.

\textbf{Control Structure Constraints:} 
We apply control flow constraints as shown in Line 6 of Algorithm~\ref{alg:test-suite-generation}, using the $getControlStmts$ function to generate a blueprint for nested control structures.
Specifically, this function first determines a random nesting depth $V$ (from 0 to a maximum $N$), then generates a sequence of $V$ control types by randomly selecting from a predefined list (e.g., [$if$, $while$, $for$]). For instance, if $V=2$, it might produce [$while$, $if$], requiring the LLM to generate code where a $while$ loop contains an $if$ statement. This ensures that the generated test case exhibits the intended control structure complexity.

\textbf{Mutation-based Augmentation:} 
To diversify the dataset, we apply a mutation step to each base test case (Line 8). Our approach builds on the type-aware mutation strategy proposed by Liu et al.~\cite{liu2023your}, but applies more thorough mutation rules to increase the difficulty. Specifically, we replace string inputs with randomly generated strings (5–20 characters) and integer inputs with values sampled within ±5 of the original. These mutations introduce variation while preserving the test’s original intent.

\textbf{Test Filtering:}
 In the second phase, each generated and mutated test case is executed. As shown in Line 13, we discard any case that results in a runtime error or produces an output longer than 50 characters. This ensures the final dataset contains only concise, valid test cases.

\begin{figure}[t]
  \centering
  \includegraphics[width=0.65\linewidth]{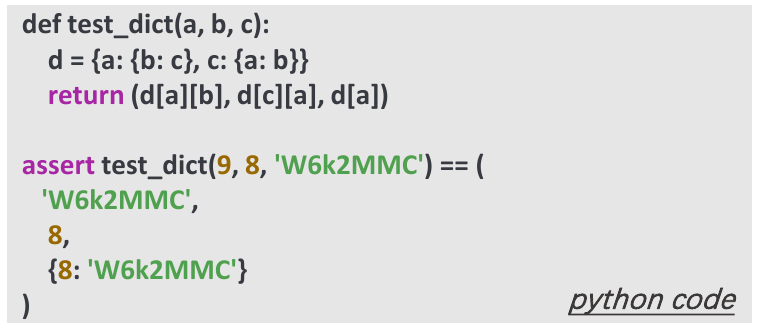}
  \caption{ A test case example from our dataset, demonstrating improved conciseness over prior work in Figure~\ref{fig:motivation2}.}\label{fig:test_case_example}
\end{figure}

In conclusion, this approach provides \textbf{fine-grained control} over the generation process through the use of constraints.
Additionally, by iterating over all methods and combining different constraints, we can generate a wide variety of test cases, ensuring the dataset is both \textbf{comprehensive} and exhibits \textbf{varied difficulty}; test cases involving more complex method interactions and control structures are naturally more challenging. 
Furthermore, because we do not require the LLM to generate a problem description from a real-world code snippet, our process naturally produces \textbf{concise} test cases free of unnecessary boilerplate.  Figure~\ref{fig:test_case_example} shows a representative test case that also covers the dict type. Compared to Figure~\ref{fig:motivation2}, this example is significantly more concise and focused on the core execution logic.

\subsection{Instruction Tuning}
Our primary objective is to enhance the LLMs' code reasoning capabilities, enabling them to accurately simulate program execution. As illustrated in Section~\ref{fig:motivation1}, small-sized LLMs often lack the knowledge about how to simulate code execution.  To address this, we first apply instruction tuning~\cite{peng2023instruction} to inject reasoning patterns into the small-sized LLMs from a powerful teacher model. This approach has proven effective in prior studies~\cite{shengyu2023instruction,ovadia2023fine,ke2025demystifying} and has also been successfully used in previous work aimed at improving code reasoning~\cite{ding2024semcoder,li2025codei}.
Moreover, this step paves the way for the subsequent reinforcement learning, since recent studies, such as Yue et al.~\cite{yue2025does}, have shown that reinforcement learning alone is ineffective at discovering entirely new reasoning paths if the LLM lacks pre-existing reasoning patterns.

This step involves two complementary tasks. The first, which we call \textit{forward reasoning}, requires the model to predict a program's output given its input. This enables our student model to learn how to accurately simulate step-by-step code execution. The second task, namely, \textit{backward reasoning}, is the inverse, i.e., the model must predict a plausible input that could produce a given output. This compels the student model to explore different execution paths more thoroughly, strengthening its overall logical reasoning capabilities.

We begin by providing the teacher model with the above tasks, which are based on the datasets constructed in Section~\ref{subsection:Dataset Construction}, and require it to generate chain-of-thought reasoning traces. 
To ensure the quality of our chain-of-thought data, we employ rejection sampling, a method used in prior works~\cite{ding2024semcoder,casella2004generalized}. For each chain-of-thought generated by the teacher model, we first extract the embedded Python code and execute it. Only if the code runs without errors is the corresponding chain-of-thought considered valid and added to our instruction tuning dataset. This process filters out flawed reasoning paths. 
Once the training data is collected, we perform instruction tuning using the base model and the validated reasoning traces.

\subsection{Reinforcement Learning} ~\label{subsection:rl}

Although instruction tuning can help LLMs acquire some level of code reasoning ability, we observe two inherent limitations. 
The first issue is poor generalization: instruction-tuned models often struggle to transfer their learned reasoning across different tasks~\cite{gupta2025selective,wang2022two}, and may even fail when presented with simple variations of the training examples~\cite{lampinen2025generalization}.
The second is the problem of overthinking or excessive reflection. The tuned model often imitates the self-reflection and self-correction patterns learned from the teacher model, but lacks the control logic to determine when to stop. This results in overly long reasoning chains or repetitive outputs~\cite{baek2025towards,yin2025towards}, especially during inference with low temperature or without a repetition penalty~\cite{keskar2019ctrl}.

To address these limitations, we apply reinforcement learning (RL) following the instruction tuning stage. Prior work~\cite{chu2025sft} has demonstrated that reinforcement learning can enhance a model's generalization ability, directly addressing the first limitation. Moreover, RL allows us to guide the model toward generating preferred outputs through a reward model. By penalizing overly long reasoning chains, we encourage the model to produce concise and efficient reasoning paths, which addresses the second limitation.
To achieve these goals, we adopt the Group-relative Policy Optimization (GRPO) algorithm~\cite{guo2025deepseek} to further refine our instruction-tuned model.
Unlike traditional methods like PPO~\cite{schulman2017proximal} that require training a separate value model, GRPO estimates advantages in a group-relative manner.
For each prompt, GRPO first generates multiple candidate responses and scores them using either a reward model or predefined rules. Finally, the model is optimized directly based on these relative advantages, avoiding the high memory cost and potential instability of training a separate value model.

Equation~\ref{eq:2} defines the group-relative advantage $\hat{A}_{i,t}$.
It quantifies the performance of a single response by normalizing its total reward $R_i$ against the mean and standard deviation of rewards from the entire group of $G$ responses.
\begin{equation}
\hat{A}_{i,t} = \frac{R_i - \text{mean}(\{R_i\}_{i=1}^G)}{\text{std}(\{R_i\}_{i=1}^G)}
\label{eq:2}
\end{equation}

To support off-policy learning, GRPO uses an importance sampling ratio  $r_{i,t}(\theta)$ given in Equation~\ref{eq:3}. It measures the probability of generating a given token $o_{i,t}$ under the new policy $\pi_{\theta}$ relative to the old policy $\theta_{\text{old}}$, enabling off-policy updates by correcting the advantage for the new policy.

\begin{equation}
r_{i,t}(\theta) = \frac{\pi_{\theta}(o_{i,t} \mid {q}, o_{i,<t})}{\pi_{\theta_{\text{old}}}(o_{i,t} \mid {q}, o_{i,<t})}
\label{eq:3}
\end{equation}

Equation~\ref{eq:4} defines the standard unclipped objective, $L_{i,t}^{\text{unclipped}}(\theta)$, which directly scales the advantage by the importance ratio. While this formulation allows the model to fully exploit advantageous updates, it can lead to unstable training. To mitigate this, 
Equation~\ref{eq:5} defines a corresponding clipped objective, $L_{i,t}^{\text{clipped}}(\theta)$, which limits the policy update by constraining the importance sampling ratio to the range $[1-\epsilon, 1+\epsilon]$. The minimum of Equations~\ref{eq:4} and~\ref{eq:5}, inspired by PPO, is a common strategy to balance effective learning with stable policy updates.

\begin{equation}
L_{i,t}^{\text{unclipped}}(\theta) = r_{i,t}(\theta)\hat{A}_{i,t} 
\label{eq:4}
\end{equation}

\begin{equation}
L_{i,t}^{\text{clipped}}(\theta) = \text{clip}(r_{i,t}(\theta), 1-\epsilon, 1+\epsilon)\hat{A}_{i,t}
\label{eq:5}
\end{equation}

Equation~\ref{eq:6} presents the final GRPO objective function $\mathcal{J}_{\text{GRPO}}(\theta)$ that is maximized during training. It integrates the previous components by first adopting the pessimistic minimum of the unclipped and clipped objectives for each token. A KL-divergence penalty is then subtracted from this value to regularize the policy. Finally, these token-level values are averaged over all tokens and responses in the batch to yield the final objective.

\begin{equation}
\label{eq:6}
\begin{aligned}
\mathcal{J}_{\text{GRPO}}(\theta) = \mathbb{E} \Biggl[ \frac{1}{G} \sum_{i=1}^G \frac{1}{|o_i|} \sum_{t=1}^{|o_i|} \biggl( & \min \left( L_{i,t}^{\text{unclipped}}(\theta), L_{i,t}^{\text{clipped}}(\theta) \right) \\
& - \beta D_{\text{KL}}(\pi_\theta \| \pi_{\text{ref}}) \biggr) \Biggr]
\end{aligned}
\end{equation}

To simplify the reward assignment, we adopt a binary scheme. We first require the LLM to structure its output using specific tags: the reasoning process within <Reasoning> tags and the final answer within <Answer> tags. The reward ${R}_{i}$ is then calculated as follows:

\begin{equation}
{R}_{i} = \begin{cases}
2.0& answer \, is \, right  \\
0.0& otherwise \\
\end{cases}
\end{equation}

The reward function assigns a positive score only when the content within the <Answer> tags is correct. During training, we impose a response length limit: any incorrect or overly long answer that exceeds this limit receives a reward of 0.0. This reward design strongly encourages the model to generate correct answers while implicitly penalizing incorrect, verbose, or repetitive outputs.
To the best of our knowledge, this is the first application of GRPO in the domain of code reasoning. Our results show that it effectively mitigates overthinking and improves generalization, outperforming both instruction-tuned and RL-only baselines.

\section{Experimental Setup} \label{sec:Experiment setup}
\subsection{Benchmark}

To evaluate our approach, we conduct experiments on three widely-used datasets designed to test a range of code reasoning abilities:

\noindent\textbf{CRUXEval \& LiveCodeBench}: The first two datasets, CRUXEval~\cite{gu2024cruxeval} and LiveCodeBench~\cite{jain2024livecodebench}, evaluate high-level execution reasoning. These datasets require the model to perform two main tasks: predicting a program's output from a given input (forward reasoning) and predicting a plausible input that produces a given output (backward reasoning).

\noindent\textbf{REval}: The third dataset, REval~\cite{chen2024reasoning}, provides a more fine-grained analysis of the model's step-by-step simulation capabilities. In addition to input-output prediction, REval requires the model to answer detailed questions about the execution trace, such as:
\begin{itemize}[leftmargin=*]
\item \textbf{Coverage Prediction}: Will a specific line of code be executed?
\item \textbf{State Tracking}: What are the type and value of a variable after a certain line is executed?
\item \textbf{Path Prediction}: Given that a line has executed, what is the next line to be executed?
\end{itemize}

For CRUXEval~\cite{gu2024cruxeval} and LiveCodeBench~\cite{jain2024livecodebench}, we follow prior work~\cite{ding2024semcoder,li2025codei} and use pass@1 as the primary evaluation metric. For REval~\cite{chen2024reasoning}, we adopt accuracy as the metric across all of its fine-grained tasks, consistent with the original benchmark's methodology. For prompt engineering, we follow the default prompts provided by each  benchmark for all experiments.

\subsection{Implementation Detail}
We synthesize the code and the reasoning path using the QwQ-32b~\cite{team2024qwq} as the teacher model. To prevent data leakage and ensure evaluation integrity, all synthesized code undergoes a rigorous decontamination process. Following the N-gram filtering method from Guo et al.~\cite{guo2024deepseek}, we discard any synthesized code that includes a 10-gram substring identical to a snippet found in our test sets. The data is excluded from all subsequent training phases, including both instruction tuning and reinforcement learning. We initially generate 20,000  cases for both supervised fine-tuning and reinforcement learning. After decontamination, we obtain 17,332 test cases for supervised finetuning and 18,796 test cases for reinforcement learning.

All experiments are conducted on a machine equipped with eight NVIDIA Tesla A800 GPUs, each with 80 GB of memory. For our study, we adopt the Qwen2.5-Coder-Instruct model~\cite{hui2024qwen2} as the primary base model, exploring different available sizes (7B, and 14B). To demonstrate the generalizability of our approach across different model architectures, we also include Llama3-Instruct-8B~\cite{touvron2023llama} as an additional base model.

During the instruction tuning stage, we train the model for three epochs with a learning rate of 1e-5, leveraging the LLaMA-Factory framework~\cite{zheng2024llamafactory}. In the subsequent reinforcement learning (RL) phase, we apply the GRPO algorithm~\cite{guo2025deepseek} using the verl framework~\cite{sheng2024hybridflow}, with a learning rate of 1e-6. During this stage, we generate five candidate responses per prompt, set a maximum response length of 4,096 tokens, and train for two epochs. 

\subsection{Baselines}

To provide a comprehensive evaluation of \approach, we compare it with both leading closed-source and open-source Large Language Models (LLMs). Specifically, we assess its performance against OpenAI's \textbf{GPT-4o}~\cite{GPT-4o} and \textbf{GPT-4o-mini}~\cite{GPT-4o-mini} to benchmark against state-of-the-art closed-source models. We also evaluate \approach against several strong open-source models, including \textbf{Qwen2.5-72B-Instruct}~\cite{team2024qwen2}, \textbf{Llama3-70B-Instruct}~\cite{touvron2023llama}, and \textbf{Qwen2.5Coder-32B-Instruct}~\cite{hui2024qwen2}. This comparison is particularly noteworthy because these open-source models have significantly larger parameter counts than \approach. It allows us to demonstrate the effectiveness of our specialized training approach in achieving competitive performance despite a smaller model size.

We also compare \approach with two additional baselines that specialize in code reasoning: \textbf{SEMCODER}\cite{ding2024semcoder} and \textbf{CODEI/O}\cite{li2025codei}. Both methods synthesize datasets by expanding real-world code snippets and requiring the LLM to complete both the code and its corresponding description. They fine-tune LLMs using chain-of-thought reasoning paths for both forward (input-to-output) and backward (output-to-input) prediction, generated by a teacher model. SEMCODER is based on DeepSeekCoder-6.7B~\cite{zhu2024deepseek} and is fine-tuned on approximately 23,000 examples. CODEI/O is built on multiple LLMs, and the base model is then fine-tuned on a significantly larger dataset of approximately 3.5 million examples. For a fair comparison, we adopt the Qwen2.5-Coder-7B version. Following prior work~\cite{chen2021evaluating,mu2024clarifygpt}, we use a temperature of 0.0 and greedy decoding for all open-source models to ensure fair and reproducible evaluation.

\section{Evaluation Results} \label{sec:Experiment results}
\subsection{How effective is \approach compared to baselines in I/O and O/I prediction?}  \label{subsec: rq1}

\begin{table}[!t]
\caption{The performance comparisons between methods in input-output prediction and output-input prediction}\label{tab:rq1}
\resizebox{0.40\textwidth}{!}{
\begin{tabular}[c]{lcccccccccc}
\toprule
\multirow{2}{*}{\textbf{Model}} & \multirow{2}{*}{\textbf{Size}}  & \multicolumn{2}{c}{\textbf{CRUXEval}}  & \multicolumn{2}{c}{\textbf{LiveCodeBench}} & \multirow{2}{*}{\textbf{Avg}}  \\ 

& & \makecell[l]{\textbf{CXEval-O}} & \makecell[r]{\textbf{CXEval-I}} & \makecell[l]{\textbf{LCB-O}} & \makecell[r]{\textbf{LCB-I}}  &  \\
\midrule

\makecell[l]{GPT-4o} & \makecell[c]{-} & \makecell[c]{{0.905}} & \makecell[c]{0.806} & \makecell[c]{{0.848}} & \makecell[c]{0.653}   & \makecell[c]{0.803}  \\

\makecell[l]{GPT-4o-mini} & \makecell[c]{-} & \makecell[c]{{0.769}} & \makecell[c]{0.673} & \makecell[c]{0.777} & \makecell[c]{0.591} & \makecell[c]{0.703}  \\
\midrule

\makecell[l]{Qwen2.5} & \makecell[c]{72B} & \makecell[c]{{0.795}} & \makecell[c]{0.746} & \makecell[c]{0.827} & \makecell[c]{0.695}  & \makecell[c]{0.766} \\

\makecell[l]{Llama 3} & \makecell[c]{70B} & \makecell[c]{{0.637}} & \makecell[c]{0.613} & \makecell[c]{0.564} & \makecell[c]{0.526}  & \makecell[c]{0.585} \\

\makecell[l]{Qwen2.5-Coder} & \makecell[c]{32B} & \makecell[c]{{0.752}} & \makecell[c]{0.834} & \makecell[c]{0.806} & \makecell[c]{0.678}  & \makecell[c]{0.768}   \\
\midrule

\makecell[l]{SEMCODER} & \makecell[c]{6.7B} & \makecell[c]{{0.625}} & \makecell[c]{0.651} & \makecell[c]{0.597} & \makecell[c]{0.530} & \makecell[c]{0.601}  \\

\makecell[l]{CODEI/O} & \makecell[c]{7B} & \makecell[c]{{0.625}} & \makecell[c]{0.679} & \makecell[c]{0.608} & \makecell[c]{0.552} & \makecell[c]{0.616}   \\

\midrule

\makecell[l]{\approach} & \makecell[c]{7B} & \makecell[c]{0.856} & \makecell[c]{{0.863}} & \makecell[c]{{0.810}} & \makecell[c]{{0.743}}  & \makecell[c]{{0.818}} \\

\makecell[l]{\approach} & \makecell[c]{14B} & \makecell[c]{\textbf{0.912}} & \makecell[c]{\textbf{0.868}} & \makecell[c]{\textbf{0.866}} & \makecell[c]{\textbf{0.825}}  & \makecell[c]{\textbf{0.868}} \\

\bottomrule
\end{tabular}

\vspace{-3mm}
}
\end{table}
Table~\ref{tab:rq1} presents the performance of \approach against baseline models on forward (-O) and backward (-I) prediction tasks in terms of pass@1. 


From the table, we observe that \approach-7B achieves performance comparable to the state-of-the-art closed-source models. It consistently outperforms GPT-4o-mini across all tasks on both datasets, with an average improvement of 16.4\% in pass@1. Compared to GPT-4o, although \approach-7B underperforms in forward prediction tasks, it significantly outperforms GPT-4o in backward prediction tasks, resulting in an overall average improvement of 3.5\%. 
Meanwhile, \approach-14B achieves the best performance across all tasks and datasets. Compared to the strongest baseline, GPT-4o, it achieves an average improvement of 8.09\%, clearly demonstrating the effectiveness and scalability of our approach.

When comparing \approach-7B to smaller-sized baselines such as SEMCODER and CODEI/O, the performance gap becomes significantly larger. \approach-7B outperforms all baselines across all tasks on both datasets. In forward prediction tasks, it surpasses the baselines by margins ranging from 40.0\% to 40.2\%, while in backward prediction tasks, the improvement ranges from 27.1\% to 40.2\%. Overall, \approach-7B achieves an average performance gain of 32.8\%.
\find{These results demonstrate the effectiveness of \approach in both forward and backward prediction tasks. It exceeds the performance of leading closed-source models like GPT-4o and all open-source baselines across all benchmarks.}

\subsection{How effective is \approach compared to baselines in more fine-grained code reasoning tasks?}  \label{subsec: rq2}

\begin{table}[!t]
\caption{The performance comparisons between methods in more fine-grained code reasoning tasks}\label{tab:rq2}
\resizebox{0.40\textwidth}{!}{
\begin{tabular}[c]{lccccccc}
\toprule
\multirow{2}{*}{\textbf{Model}} & \multirow{2}{*}{\textbf{Size}} & \multirow{2}{*}{\textbf{Coverage}}  & \multirow{2}{*}{\textbf{State}}  & \multirow{2}{*}{\textbf{Path}} & \multirow{2}{*}{\textbf{Output}} & \multirow{2}{*}{\textbf{Avg}}  \\ 

& &   & & & &  \\
\midrule

\makecell[l]{GPT-4o} & \makecell[c]{-} & \makecell[c]{{0.875}}  & \makecell[c]{{0.724}} & \makecell[c]{0.647}   & \makecell[c]{{0.845}} & \makecell[c]{{0.773}} \\

\makecell[l]{GPT-4o-mini} & \makecell[c]{-} & \makecell[c]{{0.636}}  & \makecell[c]{0.665} & \makecell[c]{0.587} & \makecell[c]{0.770} & \makecell[c]{0.636}  \\
\midrule

\makecell[l]{Qwen2.5}& \makecell[c]{72B} & \makecell[c]{{0.885}} & \makecell[c]{0.699} & \makecell[c]{0.601}  & \makecell[c]{0.836} & \makecell[c]{0.755} \\

\makecell[l]{Llama 3} & \makecell[c]{70B} & \makecell[c]{{0.853}} & \makecell[c]{0.592} & \makecell[c]{0.403}  & \makecell[c]{0.746} & \makecell[c]{0.649} \\

\makecell[l]{Qwen2.5-Coder} & \makecell[c]{32B}& \makecell[c]{{0.856}}  & \makecell[c]{0.667} & \makecell[c]{\textbf{0.687}}  & \makecell[c]{0.833}  & \makecell[c]{0.761} \\
\midrule

\makecell[l]{SEMCODER}& \makecell[c]{6.7B} & \makecell[c]{{0.467}} & \makecell[c]{-} & \makecell[c]{-} & \makecell[c]{0.562} & \makecell[c]{-} \\

\makecell[l]{CODEI/O} & \makecell[c]{7B}& \makecell[c]{{0.709}} & \makecell[c]{0.485} & \makecell[c]{0.409} & \makecell[c]{0.604} & \makecell[c]{0.552}  \\

\midrule

\makecell[l]{\approach} & \makecell[c]{7B} & \makecell[c]{0.864}  & \makecell[c]{0.672} & \makecell[c]{0.514}  & \makecell[c]{0.843} & \makecell[c]{0.723} \\

\makecell[l]{\approach} & \makecell[c]{14B} & \makecell[c]{\textbf{0.937}} & \makecell[c]{\textbf{0.799}} & \makecell[c]{0.606}  & \makecell[c]{\textbf{0.910}} & \makecell[c]{\textbf{0.811}} \\

\bottomrule
\end{tabular}

\vspace{-3mm}
}
\end{table}

From Table~\ref{tab:rq2}, we observe that \approach-7B demonstrates robust and consistent performance across the fine-grained code reasoning tasks in the REval benchmark. Compared to leading proprietary models, it decisively outperforms GPT-4o-mini across all evaluation metrics. The most notable improvements are seen in Coverage and Output prediction, where \approach-7B achieves gains ranging from 9.5\% to 39.2\%, resulting in an average improvement of 13.7\%. 
\approach-14B further raises the bar, outperforming the strongest baseline, GPT-4o, in nearly all tasks except Path prediction. On average, it achieves a 4.9\% accuracy improvement over GPT-4o, highlighting the effectiveness and scalability of our approach.

\approach-7B’s efficiency is further highlighted in comparisons with significantly larger open-source models. Despite being 5 to 10 times smaller, it outperforms Qwen2.5-72B and Qwen2.5-Coder-32B in Output prediction and consistently surpasses Llama 3-70B across all tasks. Although Qwen2.5-Coder and Qwen2.5-72B perform better on Path prediction, an area where larger  models tend to excel, \approach-7B remains highly competitive in other tasks such as Coverage and State prediction, maintaining a strong trade-off between performance and model size.

The most promising results come from direct comparisons with similarly sized models. Against SEMCODER and CODEI/O, \approach-7B holds a clear and substantial advantage. SEMCODER, in particular, fails to generate meaningful outputs in State and Path prediction which is likely due to prompt comprehension limitations and also proves the limitation of instruction tuning to apply to new tasks. \approach-7B consistently outperforms both small-scale baselines across all tasks, with relative gains ranging from 21.9\% to 39.6\%. On average, it exceeds the next-best model in its size category by 31.0\%, which underscores the strength of its training methodology and its ability to deliver high performance without relying on massive training data.

\find{
\approach delivers strong performance across fine-grained code reasoning tasks, outperforming similarly sized models by a wide margin and even surpassing leading closed-source models like GPT-4o in multiple key areas.
}

\subsection{How effective is each training stage in \approach?}  \label{subsec: rq3}

\begin{table*}[!t]
\caption{The performance comparisons in ablation study}\label{tab:rq3}
\resizebox{0.70\textwidth}{!}{
\begin{tabular}[c]{lccccccccc}
\toprule
\multirow{2}{*}{\textbf{Model}} & \multicolumn{2}{c}{\textbf{CRUXEval}}  &  \multicolumn{2}{c}{\textbf{LiveCodeBench}} & \multicolumn{4}{c}{\textbf{REval}} & \textbf{Avg}\\ 

 & \makecell[l]{\textbf{CXEval-O}} & \makecell[r]{\textbf{CXEval-I}} & \makecell[l]{\textbf{LCB-O}} & \makecell[r]{\textbf{LCB-I}} & \makecell[l]{\textbf{Coverage}} & \makecell[c]{\textbf{State}} & \makecell[c]{\textbf{Path}} & \makecell[r]{\textbf{Output}} & 
  \\
\midrule

\makecell[l]{\approach-raw} & \makecell[c]{{0.610}} & \makecell[c]{0.660} & \makecell[c]{0.580} & \makecell[c]{0.468}  & \makecell[c]{0.786} & \makecell[c]{0.517} & \makecell[c]{0.497} & \makecell[c]{0.603} & \makecell[c]{0.590}

\\

\makecell[l]{\approach-it} & \makecell[c]{{0.751}} & \makecell[c]{0.456} & \makecell[c]{0.590} &\makecell[c]{0.311} &\makecell[c]{0.851}  &\makecell[c]{0.637} &\makecell[c]{0.473} &\makecell[c]{0.814} & \makecell[c]{0.610}

\\

\makecell[l]{\approach-rl} & \makecell[c]{{0.655}} & \makecell[c]{0.680} & \makecell[c]{0.612}  & \makecell[c]{0.482} & \makecell[c]{0.772} & \makecell[c]{0.539} & \makecell[c]{\textbf{0.523}} & \makecell[c]{0.537} & \makecell[c]{0.600} 

\\

\makecell[l]{\approach-direct} & \makecell[c]{{0.538}} & \makecell[c]{{0.545}} & \makecell[c]{{0.407}} & \makecell[c]{{0.401}}  & \makecell[c]{{0.672}} & \makecell[c]{0.442} & \makecell[c]{0.477} & \makecell[c]{0.432} & \makecell[c]{{0.489}}

\\ 

\midrule

\makecell[l]{\approach} & \makecell[c]{\textbf{0.856}} & \makecell[c]{\textbf{0.863}} & \makecell[c]{\textbf{0.810}} & \makecell[c]{\textbf{0.743}} & \makecell[c]{\textbf{0.864}}  & \makecell[c]{\textbf{0.672}}& \makecell[c]{{0.514}} & \makecell[c]{\textbf{0.843}} & \makecell[c]{\textbf{0.771}} 

\\

\bottomrule
\end{tabular}

\vspace{-3mm}
}
\end{table*}

\approach is trained in two stages: instruction tuning followed by reinforcement learning. To assess the effectiveness of each training stage, we introduce three ablation variants: \approach-raw, \approach-it, and \approach-rl. \approach-raw refers to the original model without any additional training. \approach-it applies only the instruction tuning stage, while \approach-rl applies only the reinforcement learning stage, without prior instruction tuning.

Table~\ref{tab:rq3} presents the results of our ablation study. The findings indicate that both training stages, instruction tuning and reinforcement learning, are essential for the overall performance. Removing either stage leads to a significant drop in effectiveness. The \approach-it variant, which includes only instruction tuning, achieves strong results on datasets like CXEval-I and on the Coverage, State, and Output tasks in the REval benchmark. However, it also exhibits instability, performing notably worse than the untrained \approach-raw model on CXEval-O and LCB-O. This degradation is largely due to the tendency of \approach-it to produce overly long or repetitive chains of thought, an issue we will examine in more detail in Section~\ref{sec:Discussion}. 

The \approach-rl variant generally improves performance across most tasks and datasets compared to \approach-raw. However, the gains are relatively modest. While it is more stable than \approach-it, its performance still drops notably on the Output task in the REval dataset. We attribute the limited and unstable gains to the model’s lack of domain-specific knowledge in code execution, as discussed in Section~\ref{subsec:preliminary1}. This aligns with the findings of Liu et al.~\cite{liu2025understanding}, who observed that the effectiveness of reinforcement learning is constrained by the base model’s prior knowledge. When a model lacks domain expertise, reinforcement learning alone offers limited benefit. This trend is confirmed by our results: after injecting domain-specific knowledge through instruction tuning, the subsequent reinforcement learning stage leads to substantial improvements. The full \approach model outperforms both \approach-raw and \approach-it by 26.4\% to 28.5\% on average.

\begin{figure}[t]
  \centering
  \includegraphics[width=0.65\linewidth]{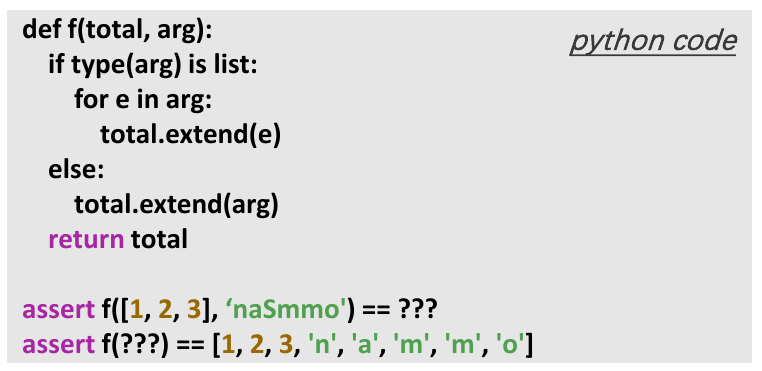}
  \caption{A case study demonstrating the effectiveness of the two-stage training }\label{fig:case_study}
\end{figure}

Figure~\ref{fig:case_study} presents a case study from the CRUXEval~\cite{gu2024cruxeval} dataset, illustrating the effectiveness of our two-stage training process. We manually examine the outputs produced by the LLM in this example. Based on the outputs, we find that \approach-raw correctly understands the function’s logic. It identifies that the function checks whether \code{arg} is a list. If it is, the function extends \code{total} with each element in the list; otherwise, it extends \code{total} with \code{arg} directly. 
However, the model fails to simulate the execution properly due to its lack of execution-related knowledge. Specifically, it does not recognize that a string is an iterable object. As a result, instead of adding each character of the string individually, it appends the entire string as a single element. This leads to an incorrect output of [1, 2, 3, 'naSmmo'], rather than the expected [1, 2, 3, 'n', 'a', 'S', 'm', 'm', 'o'].
\approach-it correctly identifies that \code{arg} is a string and treats it as an iterable of characters, producing the correct output. This demonstrates that instruction tuning enables the model to acquire domain-specific knowledge related to code execution. However, when performing backward reasoning, the model shows signs of overthinking. Its reasoning chain reveals that while it accurately recognizes \code{arg} could be a string or a list, it becomes indecisive about which type to choose when generating the input. This indecision leads to an infinite loop, reflecting the limitations discussed in Section~\ref{subsection:rl}.
After applying our full two-stage training, \approach successfully performs both forward and backward reasoning on this case.

We also introduce another variant:  \approach-direct. In this variant, the model is required to output the final answer directly, without generating any intermediate reasoning chain. As shown in Table~\ref{tab:rq3}, \approach-direct exhibits a significant drop in performance. Notably, \approach-direct performs worse than the untrained \approach-raw model. These results highlight that the generated reasoning chains play a critical role in performance gains.

\find{
The experimental results highlight the complementary roles of two training stages. Only when combined do these stages yield the full potential of the model. Furthermore, the generated reasoning chains are critical to the performance improvement.
}

\begin{table*}[!t]
\caption{The performance comparison across LLMs of different sizes and architectures}\label{tab:rq4}
\resizebox{0.80\textwidth}{!}{
\begin{tabular}[c]{lccccccccccc}
\toprule
\multirow{2}{*}{\textbf{Base Model}} & \multirow{2}{*}{\textbf{Size}} & \multirow{2}{*}{\textbf{Training}} & \multicolumn{2}{c}{\textbf{CRUXEval}}  &  \multicolumn{2}{c}{\textbf{LiveCodeBench}} & \multicolumn{4}{c}{\textbf{REval}} & \multirow{2}{*}{\textbf{Avg}}\\ 

 & & &\makecell[l]{\textbf{CXEval-O}} & \makecell[r]{\textbf{CXEval-I}} & \makecell[l]{\textbf{LCB-O}} & \makecell[r]{\textbf{LCB-I}} & \makecell[l]{\textbf{Coverage}} & \makecell[c]{\textbf{State}} & \makecell[c]{\textbf{Path}} & \makecell[r]{\textbf{Output}} & 
  \\
\midrule

\multirow{2}{*}{Qwen2.5-Coder} & \multirow{2}{*}{7B}& \makecell[c]{no} & \makecell[c]{0.610} & \makecell[c]{0.660} & \makecell[c]{0.580}  & \makecell[c]{0.468} & \makecell[c]{0.786} & \makecell[c]{0.517} & \makecell[c]{0.497} & \makecell[c]{0.603} & \makecell[c]{0.590}

\\
 & & \makecell[c]{yes} &  \makecell[c]{0.856} & \makecell[c]{0.863} & \makecell[c]{0.810}  & \makecell[c]{0.743} & \makecell[c]{0.864} & \makecell[c]{0.672} & \makecell[c]{0.514} & \makecell[c]{0.843} & \makecell[c]{\textbf{0.771($\uparrow30.7\%$)}}
\\

\multirow{2}{*}{Llama3} & \multirow{2}{*}{8B} &\makecell[c]{no} &  \makecell[c]{0.393} & \makecell[c]{0.424} & \makecell[c]{0.351}  & \makecell[c]{0.269} & \makecell[c]{0.659} & \makecell[c]{0.388} & \makecell[c]{0.284} & \makecell[c]{0.383} & \makecell[c]{0.394}

\\
 & & \makecell[c]{yes} &  \makecell[c]{0.710} & \makecell[c]{0.679} & \makecell[c]{0.574}  & \makecell[c]{0.476} & \makecell[c]{0.745} & \makecell[c]{0.129} & \makecell[c]{0.352} & \makecell[c]{0.462} & \makecell[c]{\textbf{0.516($\uparrow31.0\%$)}}
\\

\multirow{2}{*}{Qwen2.5-Coder} & \multirow{2}{*}{14B} & \makecell[c]{no} & \makecell[c]{0.796} & \makecell[c]{0.739} & \makecell[c]{0.760}  & \makecell[c]{0.582} & \makecell[c]{0.800} & \makecell[c]{0.600} & \makecell[c]{0.610} & \makecell[c]{0.819} & \makecell[c]{0.713}

\\
 & & \makecell[c]{yes} &  \makecell[c]{0.912} & \makecell[c]{0.868} & \makecell[c]{0.866}  & \makecell[c]{0.825} & \makecell[c]{0.937} & \makecell[c]{0.799} & \makecell[c]{0.606} & \makecell[c]{0.910} & \makecell[c]{\textbf{0.840($\uparrow17.8\%$)}}
\\

\bottomrule
\end{tabular}

\vspace{-3mm}
}
\end{table*}

\subsection{How effective is \approach when applied to LLMs of different sizes and architectures?}  \label{subsec: rq4}

Table~\ref{tab:rq4} presents the performance of \approach when applied to models of different architectures and sizes. For each model, we report two rows: the first shows the performance before training, and the second shows the performance after applying our full training pipeline.

From the table, we observe that all models exhibit notable performance improvements, ranging from 17.8\% to 31.1\%. As expected, larger models tend to achieve better overall performance. However, the performance gain for the 14B model is smaller compared to its smaller-sized counterparts, likely because the base model already performs strongly. When comparing across model architectures, the Qwen family consistently outperforms Llama3. This can be attributed to three main factors: (1) the Qwen2.5-7B base model is stronger than Llama3-8B, and (2) prior research suggests that Qwen models are better suited for reinforcement learning compared to Llama-based models~\cite{gandhi2025cognitive,liu2025understanding}. Researchers believe this is because Qwen models have already incorporated reasoning patterns such as self-correction and self-verification during pretraining. (3) Additionally, the Qwen2.5-Coder is specifically trained for code-related tasks.
Nevertheless, the Llama3-8B model still benefits significantly from our training approach, achieving a 31.0\% improvement on average after training.

\find{
The experimental results demonstrate that \approach can be applied to models of different sizes and architectures, with an improvement ranging from 17.8\% to 31.0\% on average.
}

\section{Discussion} \label{sec:Discussion}

\begin{figure*}[htp]
\resizebox{0.70\textwidth}{!}{
    \centering 
    \begin{subfigure}[b]{0.30\textwidth}
        \centering
        \includegraphics[width=\textwidth]{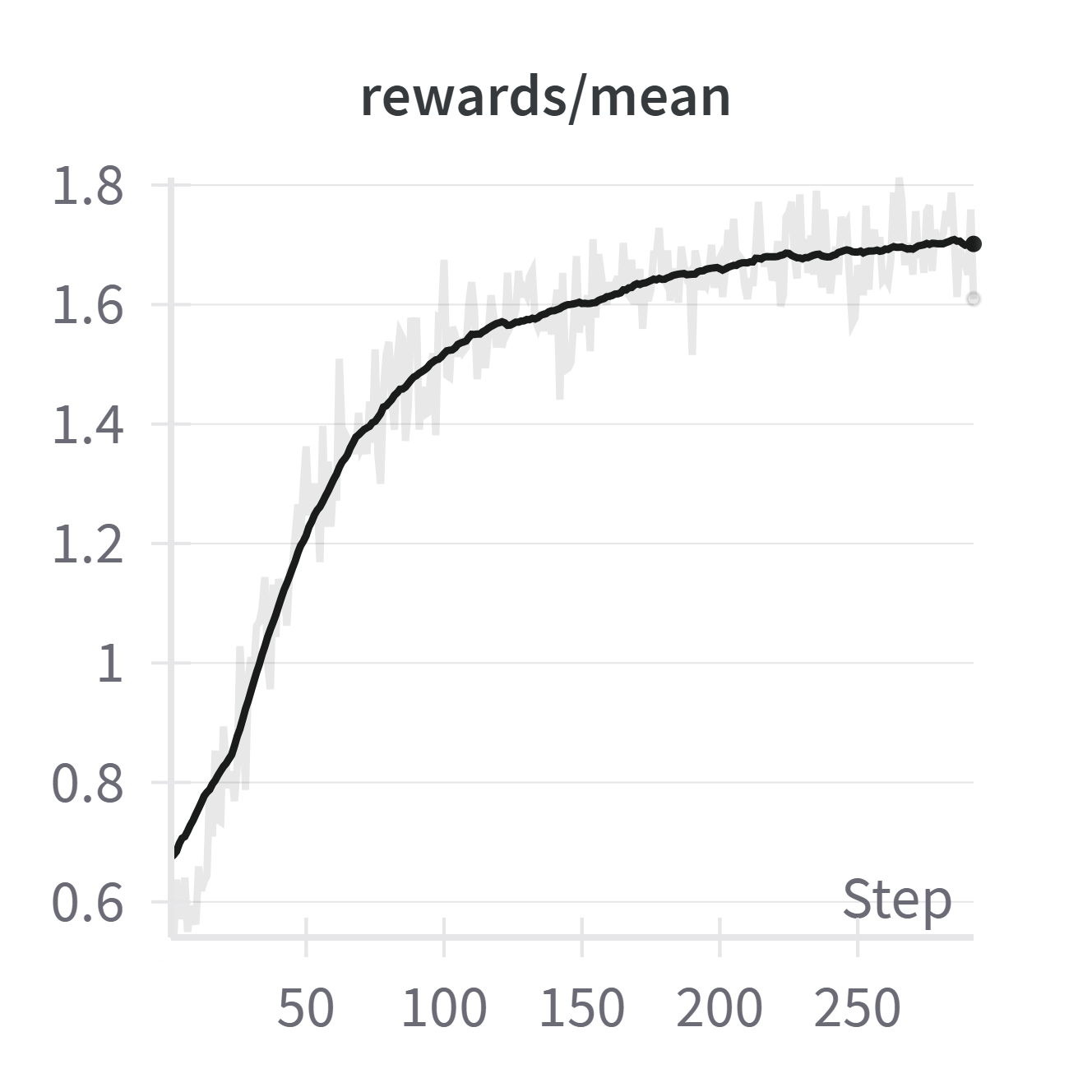}
        
        \label{fig:sub1}
    \end{subfigure}
    \hfill 
    \begin{subfigure}[b]{0.30\textwidth}
        \centering
        \includegraphics[width=\textwidth]{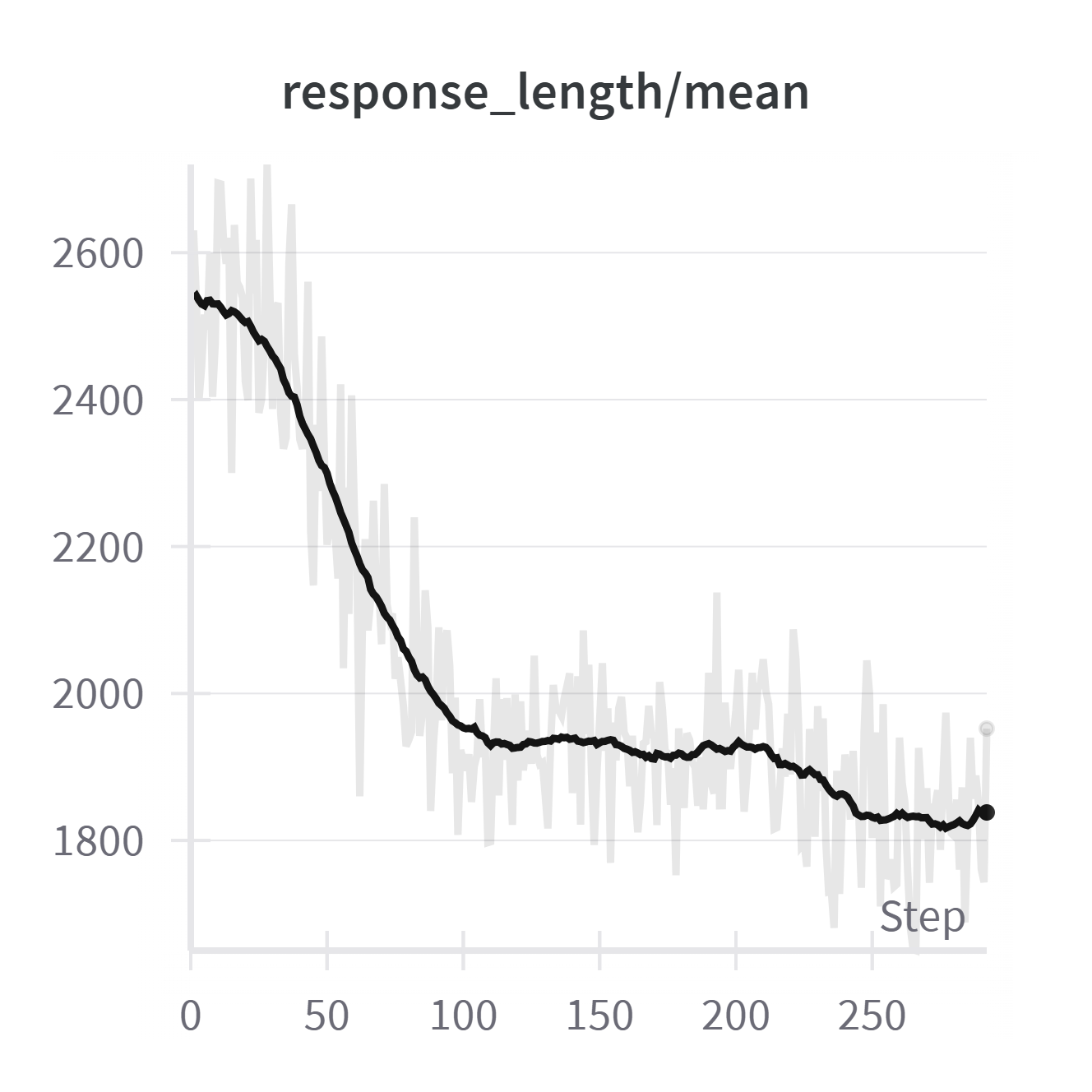}
        
        \label{fig:sub2}
    \end{subfigure}
    \hfill 
    \begin{subfigure}[b]{0.30\textwidth}
        \centering
        \includegraphics[width=\textwidth]{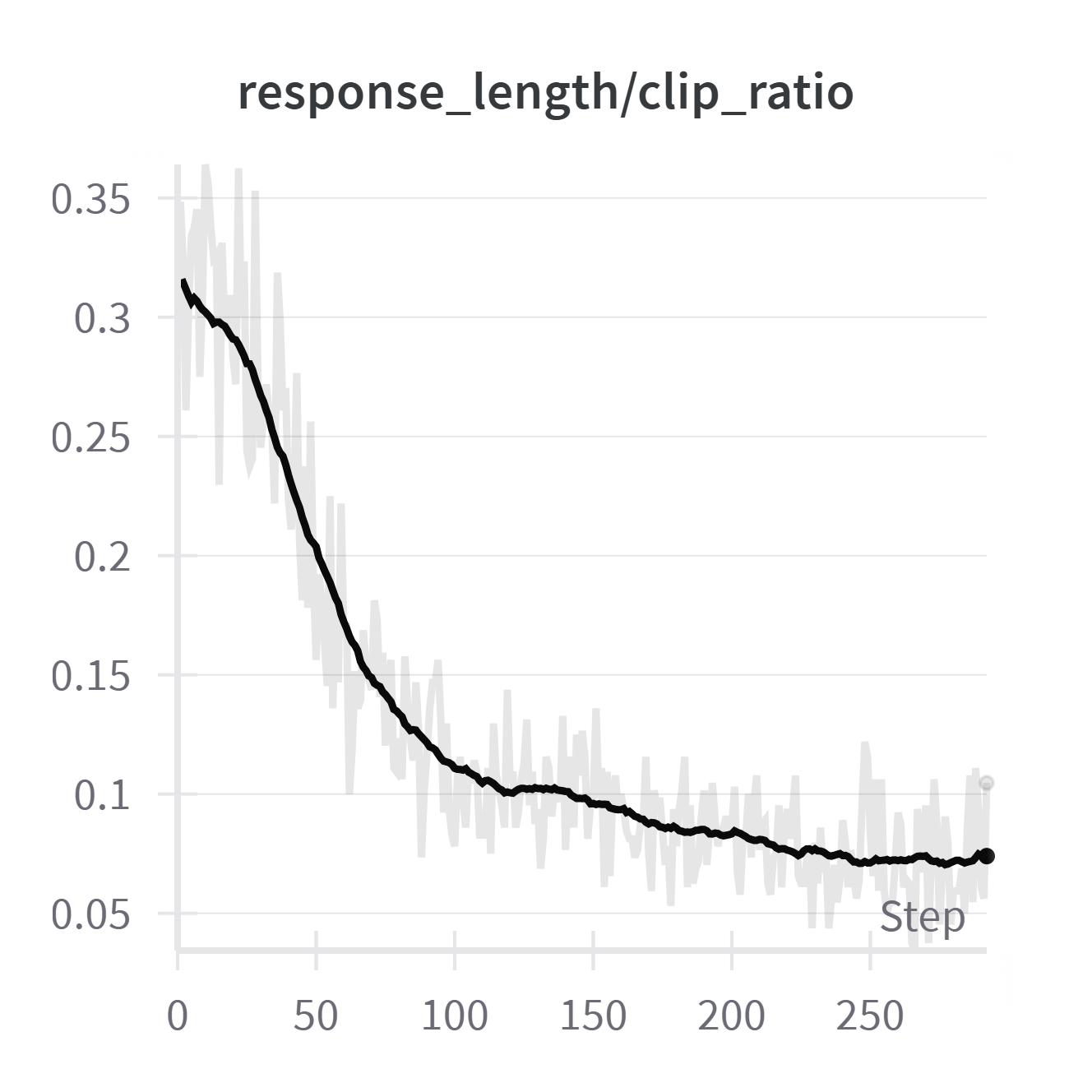}
        
        \label{fig:sub3}
    \end{subfigure}
    }
    \caption{Changes in key statistics over the training iterations}
    \label{fig:training_process}

\end{figure*}

In this section, we present key statistics from the GRPO training process, including the mean reward, the average response length, and the average clipping ratio, as shown in Figure~\ref{fig:training_process}.

From the figure, we observe that the mean reward increases steadily from 0.6 to 1.8 throughout training, indicating a stable improvement in model performance. Additionally, the clipping ratio, which is the proportion of responses that exceed the maximum allowed length and are therefore clipped, shows a significant decline over time. Initially, the clipping ratio is high, around 35\%, which aligns with our observations in Section~\ref{subsection:rl} that the model tends to generate overly long or repetitive reasoning chains after instruction tuning. As training progresses, the clipping ratio drops substantially to 5\%, suggesting that this issue is effectively mitigated. These trends highlight the effectiveness of the reinforcement learning stage in improving both the quality and efficiency of model outputs. 

One particularly interesting trend lies in the change in average response length. Prior studies~\cite{xie2025logic, zeng2025simplerl, liu2025understanding} applying reinforcement learning in domains such as mathematics often observe an increase in response length throughout training. In contrast, our training process exhibits a decrease in response length. We believe this difference can be attributed to two factors. First, it is task-specific. Not all tasks trained with GRPO lead to longer responses. For instance, Zhang et al.~\cite{zhang2025srpo} report that applying GRPO solely to code generation tasks results in shorter responses over time. Second, the model’s behavior after instruction tuning plays a key role. As discussed earlier, the model tends to produce unnecessarily long and repetitive reasoning chains. During reinforcement learning, these overly verbose responses are discouraged, and the model is incentivized to generate more concise outputs that fit within the allowed length. As a result, the proportion of excessively long responses decreases, driving down the average response length.

\section{Related Work}
\subsection{Code Reasoning}
Researchers have proposed a variety of benchmarks to evaluate large language models (LLMs) on code reasoning tasks. These range from basic input-output prediction~\cite{gu2024cruxeval,jain2024livecodebench} to more fine-grained execution analysis~\cite{chen2024reasoning}. Recently, new benchmarks have been introduced to further expand the scope of evaluation. For example, Xu et al.\cite{xu2024cruxeval} extend CRUXEval to support multiple programming languages, while Roy et al.\cite{roy2025codesense} design reasoning tasks based on real-world software projects.
Beyond execution behavior, researchers have also proposed benchmarks to assess an LLM's understanding of code semantics. Wei et al.\cite{wei2025equibench} introduce EquiBench, which requires models to determine whether two given programs are functionally equivalent. Another example is FormalBench\cite{le2025can}, in which models are asked to annotate Java programs with formal specifications. The benchmark then evaluates whether the generated specifications are logically consistent with the program and sufficiently complete in describing its behavior.

Many downstream tasks have also leveraged code execution to enhance their performance. These include program repair~\cite{ni2024next, ye2022neural}, code debugging~\cite{zhong2024debug}, code generation~\cite{ni2023lever}, and software testing~\cite{tsimpourlas2022embedding}. In addition, researchers have developed pre-trained models specifically designed for code execution~\cite{liu2023code, ding2024traced}, which have been shown to improve performance on tasks such as vulnerability detection and code clone detection.

\subsection{Reinforcement Learning in SE}
Recently, reinforcement learning has been increasingly applied in software engineering to enhance the performance of various tasks. In fuzz testing, researchers use reinforcement learning to guide fuzzers toward generating more effective inputs~\cite{li2022fuzzboost, he2024curiosity, eom2024fuzzing}. In the domain of code completion, reinforcement learning has been employed to train critic models that assess the quality of partially generated code~\cite{li2024ircoco}. For repository-level code completion, it has been used to retrieve relevant content from the codebase more effectively~\cite{wang2024rlcoder}. Reinforcement learning has also been applied to program repair~\cite{wei2025swe}, where a simple reward based on the line-level similarity between the generated and correct patches leads to significant performance improvements after training. These results highlight the broad applicability and effectiveness of reinforcement learning in software engineering tasks.

\section{Threats to Validity} \label{sec:Threats}
\noindent\textbf{Internal Validity.} The teacher LLM may occasionally generate incorrect code when constructing the dataset or produce flawed reasoning chains during forward and backward prediction. To mitigate risks to dataset quality, we execute all test cases generated by the teacher model and retain only those that are runnable. To ensure the quality of the reasoning chains, we further validate them by running the associated test cases and keeping only the chains that lead to the correct answer. While it is possible that some reasoning chains may be logically incorrect despite producing the correct output, we believe such cases are very rare given that we use a highly capable teacher model (Qwen-32B) and therefore have a minimal impact on overall dataset quality.

\noindent\textbf{External Validity.}
External validity refers to generalizability of our approach. One common concern is whether \approach can be effectively applied to other large language models. To address this, we apply our dataset and two-stage training pipeline to LLMs of different architectures and sizes. Due to computational constraints, we currently evaluate our method on 7B and 14B models. Nevertheless, the strong performance observed in these settings suggests that our approach generalizes well beyond a single model family or size. Another potential threat to generalizability is that our current implementation focuses solely on Python. However, Python is one of the most popular programming language. To further strengthen the external validity of our work, we also plan to extend \approach to support additional programming languages in future research. Additionally, we also aim to test our method on larger-scale models to fully explore its scalability.

\section{Conclusion and Future Work}
In this paper, we propose \approach, a novel technique that spans from training dataset construction to a two-phase training framework. During dataset construction, we focus on capturing the core logic of code execution while eliminating irrelevant content such as boilerplate code. In the training framework, we first inject code reasoning knowledge into the LLM through instruction tuning. Then, we apply reinforcement learning to further enhance the model's performance and generalization capabilities. Through extensive evaluation across multiple datasets, \approach demonstrates substantial improvements over small-sized baselines and achieves performance comparable to advanced models like GPT-4o on most tasks. In future work, we plan to extend our dataset to include additional programming languages and apply \approach to multilingual code reasoning benchmarks.  We also aim to utilize \approach as a foundation for intelligent developer tools, such as debugging and repair assistants.

\bibliographystyle{ACM-Reference-Format}
\bibliography{references.bib}

\end{document}